# CASTOR status and evolution

Jean-Philippe Baud, Ben Couturier, Charles Curran, Jean-Damien Durand, Emil Knezo, Stefano Occhetti, Olof Bärring
*CERN, CH-1211 Geneva-23, Switzerland*

In January 1999, CERN began to develop CASTOR ("CERN Advanced STORage manager"). This Hierarchical Storage Manager targetted at HEP applications has been in full production at CERN since May 2001. It now contains more than two Petabyte of data in roughly 9 million files. In 2002, 350 Terabytes of data were stored for COMPASS at 45 MB/s and a Data Challenge was run for ALICE in preparation for the LHC startup in 2007 and sustained a data transfer to tape of 300 MB/s for one week (180 TB). The major functionality improvements were the support for files larger than 2 GB (in collaboration with IN2P3) and the development of Grid interfaces to CASTOR: GridFTP and SRM ("Storage Resource Manager"). An ongoing effort is taking place to copy the existing data from obsolete media like 9940 A to better cost effective offerings. CASTOR has also been deployed at several HEP sites with little effort. In 2003, we plan to continue working on Grid interfaces and to improve performance not only for Central Data Recording but also for Data Analysis applications where thousands of processes possibly access the same hot data. This could imply the selection of another filesystem or the use of replication (hardware or software).

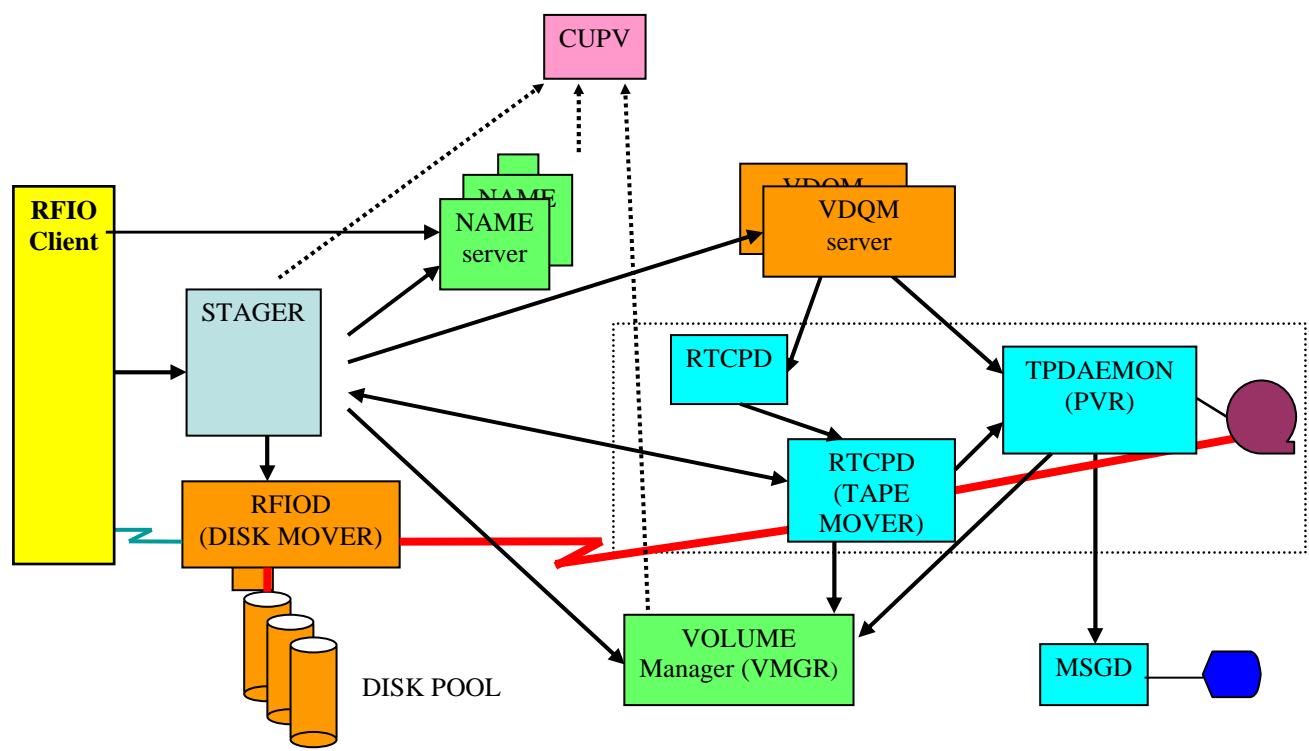

**Figure 1: CASTOR component architecture**

## 1. INTRODUCTION

CASTOR stands for CERN Advanced STORage manager. It is a hierarchical storage management system designed for (but not constrained to) storing HEP users' and experiments' files. It manages both secondary (disk) and tertiary (tape) storage. The CASTOR software development started in 1999 and was put in full production in 2001. The software is an evolution of SHIFT, which was the tape and disk management system developed and used by CERN since the beginning of 1990s. CASTOR manages today over 9 million files and 2PetaBytes of data.

The next section describes the software architecture and the main characteristics of the CASTOR software.

Thereafter some statistics for its current use at CERN are presented.

## 2. CASTOR ARCHITECTURE AND CHARACTERISTICS

The CASTOR software architecture is schematically depicted in Figure 1. The client application should normally use the Remote File IO (RFIO) library to interface to CASTOR. RFIO provides a POSIX compliant file IO and metadata interface, e.g. *rfio_open(), rfio_read(), rfio_lseek(), rfio_stat()*, etc. The Name server provides the CASTOR namespace, which appears as a normal UNIX filesystem directory hierarchy. The name space is rooted with "/castor" followed by a directory that specifies the domain, e.g. "/castor/cern.ch" or "/castor/cnaf.infn.it". The next level in the name space hierarchy identifies the hostname (or alias) for a node with a running instance of the CASTOR name server. The convention is *nodename = "cns" + "directory name"*, e.g. "/castor/cern.ch/user" points to the CASTOR name server instance running on a node *cnsuser.cern.ch*. The naming of all directories below the third level hierarchy is entirely up to the user/administrator managing the sub-trees.

The RFIO client library interfaces to three CASTOR components: the name server providing the CASTOR namespace described in previous paragraph; the CASTOR stager, which provides the disk pool management for space allocation and recall/migration of tape files; the RFIO server (rfiod), which is a multithreaded daemon responsible for the physical I/O between the client and the disk file.

All tape access is managed by the CASTOR stager. The client does not normally know about the tape location of the files being accessed. The stager therefore interfaces with:

- The CASTOR Volume Manager (VMGR) to know the status of tapes and select a tape for migration if the client created a new file or updated an existing one
- The CASTOR Volume and Drive Queue Manager (VDQM), which provides a FIFO queue for accessing the tape drives
- The CASTOR tape mover, Remote Tape COPY (RTCOPY), which is a multithreaded application with large memory buffers, for performing the copy between tape and disk
- The RFIO server (rfiod) for managing the disk pools

CASTOR is a modular software system where each component has a well-defined role. This allows changing components without affecting the whole system.

The system also allows for very distributed configurations. The number of disk and tape servers is not limited by software.

In order to assure scalability, centralized services like the CASTOR name server can be distributed. The CASTOR name server and VMGR have RDBMS back-ends where currently Oracle (used at CERN) and MySQL are supported.

The CASTOR software has been compiled and tested on a large variety of hardware: Linux, Solaris, AIX, HP-UX, Digital UNIX, IRIX and Windows (NT and W2K). The CASTOR tape software supports DLT/SDLT, LTO, IBM 3590, STK 9840, STK9940A/B tape drives[1] and ADIC Scalar, IBM 3494, IBM 3584, Odetics, Sony DMS24, STK Powderhorn tape libraries as well as all generic SCSI driver compatible robotics.

## 3. CASTOR AND THE GRID

The GridFTP server from the Globus 2.2 Toolkit has been ported to RFIO. This is now being set up as a production service allowing experiments to import/export data from CERN.

A prototype server partially implementing the Storage Resource Manager (SRM) version 1.0 definition has been set up and was demonstrated to work in basic interoperability tests with an SRM client running at FNAL.

## 4. CASTOR AT CERN

CASTOR manages almost all physics data at CERN. The current setup for the production service includes about 80 tape drives distributed over 60 tape servers and about 40 stagers managing the space on some 140 disk servers.

| *Model* | *Nb Drives* | *Nb Servers* |
|---|---|---|
| 9940B | 21 | 20 |
| 9940A | 28 | 10 |
| 9840 | 15 | 5 |
| 3590 | 4 | 2 |
| DLT7000 | 6 | 2 |
| LTO | 6 | 3 |
| SDLT | 2 | 1 |

**Table 1: tape drives used at CERN**

Table 1 summarizes the tape drives being used at CERN. The tape library infrastructure consists of 2 geographically separated installations with 5 Powderhorn silos each providing a total slot space for 2x27500 cartridges.[2]

CASTOR manages about 9 million files with a total data volume of more than 2PetaByte. Despite the large number of disk servers, tape is a very active storage media.

## 5. CONCLUSIONS

CASTOR is a scalable and distributed hierarchical storage management system. It is used in production at CERN since 2001. The system has also been installed and is running for production use in other sites, e.g. CNAF (Bologna) in Italy. CASTOR differs from several other HSM systems in that it not only provides a tape archive with a logical namespace but also a full-fledged disk management layer.

---

[1] Also a large number of old tape drives like 3480 and 3490 are supported.

[2] The tape drives numbers reflect the installation in March 2003. Since then all 9940A model drives have been replaced with 9940B model for a total of 50 9940B drives.